# USAGE OF POROUS Al$_2$O$_3$ LAYERS FOR RH SENSING

*Veronika Timár-Horváth, László Juhász, András Vass-Várnai, Gergely Perlaky*
timarne@eet.bme.hu, blue@impulzus.com, a.v.varnai@gmail.com, perlaky@eet.bme.hu

Budapest University of Technology and Economics (BUTE), Department of Electron Devices,
Budapest, Hungary

**ABSTRACT**

At the Department of Electron Devices a cheap, more or less CMOS process compatible capacitive type RH sensor has been developed. Capacitive sensors are based on dielectric property changes of thin films upon water vapor uptake which depends on the surrounding media's relative humidity content. Because of the immense surface-to-volume ratio and the abundant void fraction, very high sensitivities can be obtained with porous ceramics. One of the ceramics to be used is porous Al$_2$O$_3$, obtained by electrochemical oxidation of aluminum under anodic bias. The average pore sizes are between 6…9 nm. In our paper we intend to demonstrate images representing the influence of the technological parameters on the porous structure and the device sensitivity.

## 1. INTRODUCTION

Humidity is an increasingly important factor, which affects industrial, agricultural processes and influences even our everyday-life. For this reason, at the Department of Electron Devices a low-cost, capacitive RH sensor has been developed using CMOS process compatible technology and an additional anodization step.

Capacitive sensors are based on dielectric property changes of thick or thin films upon water vapor uptake which depends on the surrounding media's relative humidity content. Due to the polar structure of the H$_2$O molecule, water exhibits a very high permittivity $\varepsilon_w$=80 at room temperature. The permittivity of dielectric films shows a huge increase upon adsorption of water. In a porous dielectric the air in the voids is replaced by adsorbed vapor as the ambient humidity level increases therefore it's worth using porous dielectrics instead of compact layers. [1]

Capacitive sensors – as well as other absorption-based humidity sensors – typically show a non-linear behavior as function of RH. This behavior can be described by the following phenomenological equation:

$$\frac{C_w}{C_d} = \left(\frac{\varepsilon_w}{\varepsilon_d}\right)^n, \qquad (1)$$

where $\varepsilon_d$ and $\varepsilon_w$ are the permittivity of the dielectric material at dry and wet state, $C_w$ and $C_d$ are the corresponding capacitances and $n$ is a factor related to (the morphology of) the dielectric film [1].

Because of the immense surface-to-volume ratio and the abundant void fraction, very high sensitivities can be obtained with porous ceramics. One of the ceramics to be used is porous Al$_2$O$_3$ which has proven to be stable at elevated temperatures and at high humidity levels. A well established method for porous Al$_2$O$_3$ preparation is the electrochemical oxidation of Al films under anodic bias.

The quality of the Al$_2$O$_3$ layers and consequently the sensor behavior (sensitivity, humidity range, response time, stability, etc.) strongly depend on the thickness of the porous layer and the density and size of the pores as well [2]. The structure of porous alumina prepared by the aforementioned method can be influenced by the technological parameters such as concentration and temperature of the electrolyte and the current density in the anodizing cell [3] [4] [5].

Theoretically, the best solution to cover the total humidity range (0…100%) is to produce all size of pores: nano-, mezo- and macropores. According to a recent study [4] a highly reproducible, wide-range humidity sensor can be achieved using nanodimensional pores of a narrow size distribution.

The response time of ceramic humidity sensors is generally limited by diffusion [6]. Moreover, ceramics are highly sensitive to contaminants such as dust and smoke. These sensors require maintenance: having the condensed vapor evaporated by heating up them from time to time. The temporary heating could also be a solution to reduce the drift due to the formation of chemisorbed OH$^-$ groups [4] [7].

## 2. EXPERIMENTAL

Highly n-doped 2" silicon wafers were used as substrate. Wafers were cleaned in RBS detergent, refreshed in dilute (1:20) HF, boiled in d.i. water and after drying oxidized in 100 l/hour dry O$_2$ flow at 1100 °C for 20 or 50 minutes.

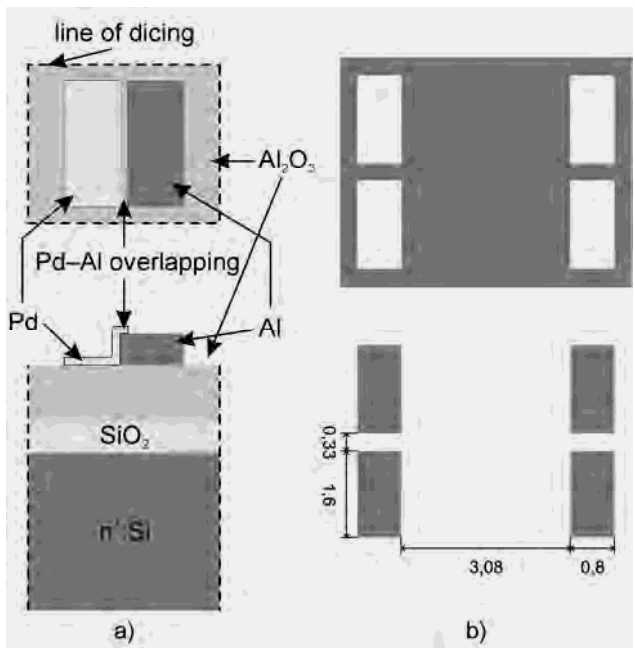

*Fig. 1. Structure of the capacitive humidity sensor (a) and part of the masks used for photolithography (b). (All sizes are given in mm.)*

This step resulted in a 70 nm or 117 nm thick SiO$_2$ layer serving as insulation [8] and adhesive layer. Vacuum vapor deposition was carried out to create the aluminum thin film layer over the SiO$_2$ using 99.99% purity wire as source material. The aluminum was anodized in aqueous solution of 2.7 mol/l (23.1 wt%) sulphuric acid [3]. The anodic current density was kept at a constant value (5…20 mA/cm$^2$) [3], which fell suddenly to zero when the metal became completely oxidized. Another aluminum layer was deposited on the top of the sensing layer and usual photolithographic steps were used to form contact pads from it. The structure was spin-coated with photoresist, which was patterned using the inverse of the previous mask shifted slightly, to result overlapping windows with the aluminum pads. A 12 nm thin layer of palladium was RF-sputtered onto the surface and it was machined in a "lift-off" manner with the prepared photoresist (Fig. 1.). The ready-made wafer was cut to dices and the selected dices were mounted onto TO-5 type transistor headers with eutectic bond. The aluminum pad of each chip was connected to a lead with thermo-compression using 50 μm gold wire (Fig. 2.).

Starting with the same substrate material, aluminum and SiO$_2$ layers with different thickness were tested to investigate the influence of these parameters on the sensor's properties. For the same reason different current densities and bath temperatures were used during anodic oxidation (Table I., Fig. 3.).

*Table I. Anodization conditions used for making porous alumina layers.*

| Wafer | SiO$_2$ [nm] | Al [nm] | Al heat-treated | J$_{ox}$ [mA/cm$^2$] | T$_{ox}$ [°C] |
|---|---|---|---|---|---|
| #4 | 70 | 440 | – | 10 | 14 |
| #5 | 70 | 440 | – | 10 | 16 |
| #6 | 70 | 440 | – | 10 | 18 |
| #15 | 117 | 20 | + | 3.75 | 16 |
| #11 | 70 | 30 | + | 7.5 | 18 |
| #20 | 70 | 55 | – | 3.75 | 16 |
| #13 | 70 | 65 | + | 5 | 18 |
| #21 | 117 | 75 | – | 5 | 16 |

### 3. RESULTS AND DISCUSSION

For testing the devices, two types of test environments were used: fixed-point humidity environments over five different saturated salt solutions at 25 °C [9], and a humidity chamber (ESPEC SH-241). Capacitance measurements were performed using HAMEG LC meter at 16 kHz. Typical sensing characteristics are shown in Fig. 4.

Both the characteristics measured over saturated salt solutions and in the humidity chamber show the same sensitivity of our structure, where the average sensitivity is about 5 pF/RH%. Although the results are very promising, there are still some side effects that have to be taken into account.

On one hand the sensing characteristics aren't linear, as desired, moreover there is a large hysteresis loop (type IV) [10] [11] in the sensors capacitance–RH characteristics, on the other hand there is a temperature dependence of the vapor adsorption phenomena in the porous structure.

### 3.1. Hysteresis loop and linearity

According to the operation theory of the capacitive RH sensors water molecules are adsorbed onto and desorbed from the wall of the pores as a function of the ambient humidity content of the gas and the surface properties of the ceramics. According to the theory of porous dielectric AAO (anodized aluminum oxide) sensors, the wetting phenomena have two major domains [10].

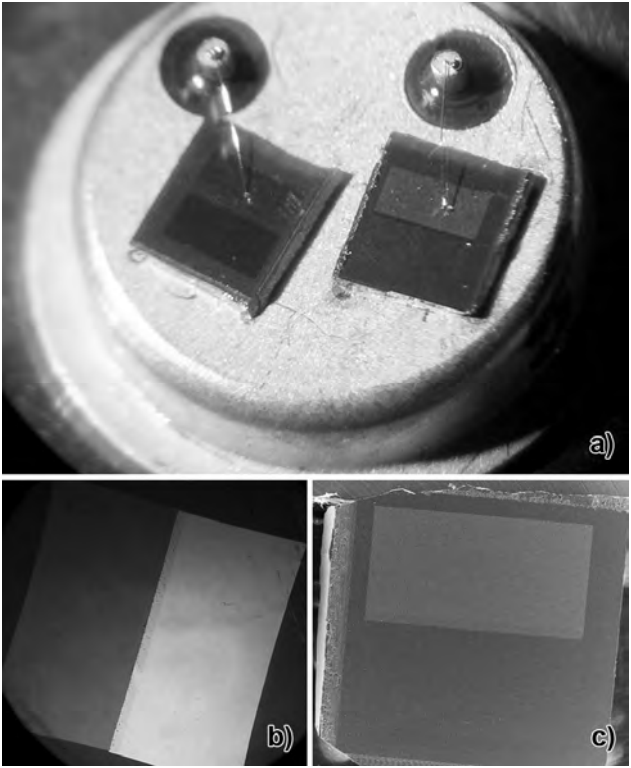

*Fig. 2. Digital camera photo of two sensors mounted on the same TO-5 type header (a), microscopic view of the structure on wafer (b) and SEM image of a chip (c).*

At relative low RH levels, very thin layers develop from water molecules on the pore walls and heat of condensation releases. This phenomenon can be explained by the BET theory of adsorption on nanoporous materials [12] [10]. The BET theory assumes that the solid material has fixed number of adsorption sites and the adsorbed layers can be several molecules thick. It is also put forward, that the heat of adsorption in all layers beyond the first is equal to the latent heat of condensation

$$\frac{p}{v(p_0-p)} = \frac{1}{v_m c} + \frac{c-1}{v_m c}\frac{p}{p_0}, \qquad (2)$$

where $p$ is pressure of the gas, $p_0$ is the saturated vapor pressure, $v$ is the quantity of the adsorbed gas, $v_m$ is the monolayer capacity of the surface (the quantity of one adsorbed layer), $c$ is a temperature dependent value:

$$c = e^{\frac{E_1-E_L}{RT}}, \qquad (3)$$

where $E_1$ is the adsorption heat of the first layer and $E_L$ is the heat of condensation. A further assumption is that the vapor condenses on the adsorbent as a liquid when its pressure reaches the saturated vapor pressure, i.e., at $p=p_0$ the number of the adsorbed layers is infinite.

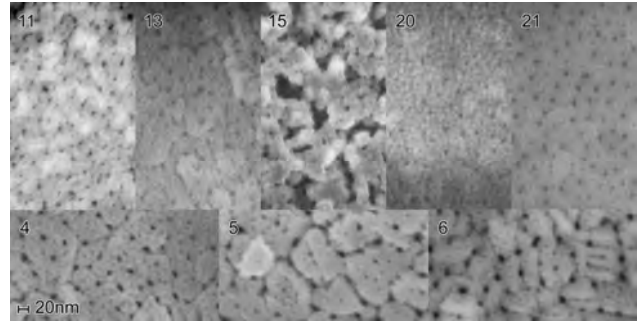

*Fig. 3. SEM images of the different wafers. The average pore diameters are in the range of 6…9 nm.*

In a narrow pored adsorbent, instead of infinite layers of water molecules, there's room only for $n$ layers. The BET equation for $n$ adsorbed layers shows:

$$\frac{v}{v_m} = \frac{c \cdot (p/p_0)}{1-p/p_0} \cdot \frac{1-(n+1)(p/p_0)^n + n(p/p_0)^{n+1}}{1+(c-1)(p/p_0)-c(p/p_0)^{n+1}}. \quad (4)$$

This equation can describe the adsorption branch of the type IV isotherm (see Fig. 4.).

As the humidity level increases, the pores start to fill. This filling can be explained by the capillary condensation theory. In case of higher RH it has been shown by Kelvin, that at a given pressure vapor condenses into those pores, which have radii smaller than the Kelvin radius, given by following equation:

$$\ln(p/p_0) = -\frac{2\gamma V}{rRT} \cdot \cos\theta, \qquad (5)$$

where $\gamma$ is the surface tension of the liquid, $V$ is the molecular volume and $R$ and $T$ have their usual meanings. The negative sign implies for $\theta<90°$, i.e. $p$ is less than $p_0$.

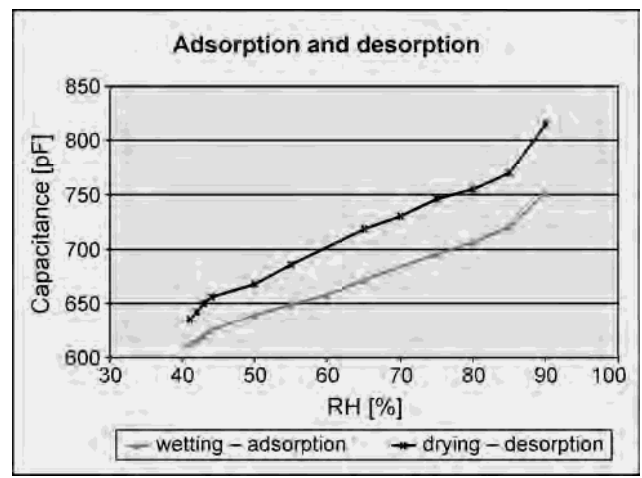

*Fig. 4. Typical RH–capacitance characteristics of the produced RH sensor*

The capillary condensation might cause hystheresis in the characteristics [10]. As *p* increases, wider and wider pores will be filled, as *p* decreases, the pores get empty. There are two different contact angles on the two branches of the loop, while the pores are filling, there is an advancing contact angle, $\theta_A$, and while they are emptying, there is a "receding" contact angle, $\theta_R$. Since $\theta_A$ is greater than $\theta_R$, according to the equation (5), the value of *p* for a given value of *r* is less during desorption than adsorption causing the hysteresis (Fig. 4.). This gives the explanation of the upper part of the sensing curve (desorption).

Over 80% RH level the formation of a monomolecular water layer might start on the entire surface causing a sensitivity enhancement.

### 3.2. The role of ambient pressure

Measurements were carried out at 2 atm, 90% RH. Under these conditions the sensors' capacitance became approximately 5 times higher than at the same RH at normal pressure. This means the pores of the produced sensing layer cannot be filled completely at atmospheric pressure even at high humidity.

### 3.3. Temperature dependence

The other major side effect is the temperature dependence of the vapor adsorption phenomena in the porous structure, which has been proven by measurements. A constant RH was maintained in the humidity chamber and the temperature was varied in the range of 5…95 °C. Prior to the measurements the sensors were heated up to 100 °C and kept at this temperature for 10 minutes resulting in dry pores. The results are shown in Fig. 5.

A possible explanation of this phenomenon was given by R.K.Nahar [6]. He suggested that besides the capillary condensation of water vapor, there is lateral moisture diffusion into the pore walls as well, causing a virtual widening of pores especially at higher temperatures, following the exponential diffusion rules. The T–C curve's hystheresis might be caused by the difference between the diffusion mechanism of the water molecules penetrating into or exiting from the structure of the pore walls or the grain boundaries of the alumina layer. This phenomenon might affect the original sensing mechanism as well.

### 4. CONCLUSIONS

The AAO sensors are proven to be highly sensitive for ambient RH changes. In our structure the average sensitivity was ~5 pF/RH%. The shape of characteristics is not linear according to the BET theory. Above 80 RH% an increased sensitivity was observed.

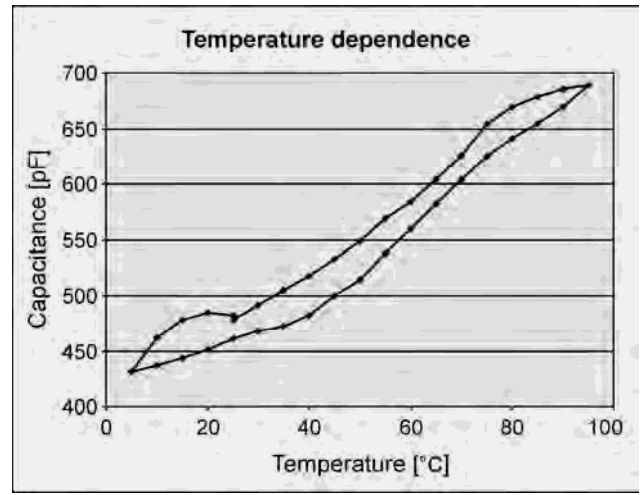

*Fig. 5. The measured temperature–capacitance characteristics at constant 35% RH*

The hysteresis in the sensing characteristics can be explained by physical laws like adsorption–desorption and capillary condensation phenomena. The nanosized structure requires the development of new, highly sensitive measuring methods in order to find the relevant explanation of the above mentioned phenomena. To overcome the parasitic temperature dependence of the AAO RH sensors electronical signal processing may be used.

Our investigations did not show significant dependence of the varied anodizing parameters on the pore size distribution, but the sensitivity depends on the initial aluminum layer thickness.

The processing of sensors has a slight difference compared to CMOS technology. The only non-standard technology step is the electrochemical oxidation of a thin aluminum film under anodic bias. The solution can be the same as for MEMS: either post-CMOS processing or chip-on-board technology. This way the integration of read-out circuitry, a microheater for conditioning the sensing layer and a transceiver would be possible.

Based on the current results and experience further steps should be taken to determine the optimal technological parameters such as layer thicknesses, composition and concentration of solution, anodizing voltage or current density. It's also necessary to select an appropriate, corrosion-resistant pad-material and overcome the packaging difficulties. CMOS circuits will need protection, while the sensor has to be exposed to the environment.

## 5. ACKNOWLEDGEMENTS

The present research is a part of the development of a smart sensor and read-out circuitry for environment monitoring by employing new materials and semiconductor technology. It is partly funded by the project PATENT DfMM in FP6 of the European Union.

## 6. REFERENCES

[1] Z. M. Rittersma: "Recent achievements in miniaturized humidity sensors – a review of transduction techniques", *Sensors and Actuators*, A 96, 2002, pp. 196-210.

[2] Goutam Banerjee, Kamalendu Sengupta: "Pore size optimisation of humidity sensor – a probabilistic approach", *Sensors and Actuators*, B vol. 86., 2002, pp 34-41.

[3] Losonci Iván, Pető Csaba, Tihanyi Kálmán: "Galvanotechnikai zsebkönyv", (Pocket-book of galvanotechnology – in Hungarian) Műszaki Kiadó, Budapest, 1982, pp. 395

[4] Oomman K. Varghese, Dawei Gong, Maggie Paulose, and Keat G. Ong, Craig A. Grimes, Elizabeth C. Dickey: "Highly ordered nanoporous alumina films: Effect of pore size and uniformity on sensing performance", *J. Mater. Res.*, Vol. 17, No. 5, 2002, pp. 1162-1171.

[5] O. Jessensky, F. Müller, U. Gösele: "Self-organized formation of hexagonal pore arrays in anodic alumina"; *Applied Physics Letters*, Vol. 72., Num.10., 1998, pp. 1173–1175

[6] R.K. Nahar: "Study of the performance degradation of thin film aluminum oxide sensor at high humidity", *Sensors and Actuators*, B 63, 2000, pp. 49–54.

[7] Ching-Liang Dai: "A capacitive humidity sensor integrated with micro heater and ring oscillator circuit fabricated by CMOS-MEMS technique", *Sensors and Actuators* B, 2006

[8] Giorgio Sberveglieri, Roberto Murri, Nicola Pinto: "Characterization of porous $Al_2O_3$-SiO2/Si sensor for low and medium humidity ranges", *Sensors and Actuators* B 23, 1995, pp. 177–180

[9] T. Lu, C. Chen: "Uncertainty evaluation of humidity sensors calibrated by saturated salt solutions"; *Measurement*, 2006, doi:10.1016/j.measurement.2006.09.012

[10] S.J. Gregg, "The Surface Chemistry of Solids", Chapman & Hall Ltd., 1961, pp. 28-82.

[11] C. Sangwichien, G.L. Aranovich, M.D. Donohue: "Density functional theory of adsorption isotherms with hysteresis loops", *Colloids and Surfaces A: Physicochemical and Engineering Aspects*, Vol. 206, No. 1-3, 2002, pp. 313-320.

[12] Stephen Brunauer, P. H. Emmett, Edward Teller: "Adsorption of Gases in Multimolecular Layers", *J. Am. Chem. Soc.*, 1938, 60, 309. doi:10.1021/ja01269a023